\documentclass[
    reprint,
    superscriptaddress,
    floatfix,
    preprintnumbers,
    longbibliography,
    prd,
    amsmath,
    amssymb,
    aps,
    nofootinbib
]{revtex4-2}

\usepackage[caption=false]{subfig}
\usepackage{graphicx}
\usepackage{dcolumn}
\usepackage{bm}
\usepackage{hyperref}
\usepackage{units}
\usepackage[dvipsnames]{xcolor}

\newcommand{\lowMass}{\textcolor{black}{\ensuremath{\unit[110]{MeV}}}}
\newcommand{\highMass}{\textcolor{black}{\ensuremath{\unit[155]{MeV}}}}

\newcommand{\fhcPot}{\textcolor{black}{\ensuremath{8.97\times 10^{20}}}}
\newcommand{\rhcPot}{\textcolor{black}{\ensuremath{1.12\times 10^{21}}}}
\newcommand{\totalPot}{\textcolor{black}{\ensuremath{2.01\times 10^{21}}}}

\begin{document}

\preprint{FERMILAB-PUB-25-0012-PPD}

\widetext

\title{Search for the production of Higgs-portal scalar bosons in the NuMI beam using the MicroBooNE detector}%

%


\newcommand{\ANL}{Argonne National Laboratory (ANL), Lemont, IL, 60439, USA}
\newcommand{\Bern}{Universit{\"a}t Bern, Bern CH-3012, Switzerland}
\newcommand{\BNL}{Brookhaven National Laboratory (BNL), Upton, NY, 11973, USA}
\newcommand{\UCSB}{University of California, Santa Barbara, CA, 93106, USA}
\newcommand{\Cambridge}{University of Cambridge, Cambridge CB3 0HE, United Kingdom}
\newcommand{\CIEMAT}{Centro de Investigaciones Energ\'{e}ticas, Medioambientales y Tecnol\'{o}gicas (CIEMAT), Madrid E-28040, Spain}
\newcommand{\Chicago}{University of Chicago, Chicago, IL, 60637, USA}
\newcommand{\Cincinnati}{University of Cincinnati, Cincinnati, OH, 45221, USA}
\newcommand{\CSU}{Colorado State University, Fort Collins, CO, 80523, USA}
\newcommand{\Columbia}{Columbia University, New York, NY, 10027, USA}
\newcommand{\Edinburgh}{University of Edinburgh, Edinburgh EH9 3FD, United Kingdom}
\newcommand{\FNAL}{Fermi National Accelerator Laboratory (FNAL), Batavia, IL 60510, USA}
\newcommand{\Granada}{Universidad de Granada, Granada E-18071, Spain}
\newcommand{\IIT}{Illinois Institute of Technology (IIT), Chicago, IL 60616, USA}
\newcommand{\ICL}{Imperial College London, London SW7 2AZ, United Kingdom}
\newcommand{\Indiana}{Indiana University, Bloomington, IN 47405, USA}
\newcommand{\KSU}{Kansas State University (KSU), Manhattan, KS, 66506, USA}
\newcommand{\Lancaster}{Lancaster University, Lancaster LA1 4YW, United Kingdom}
\newcommand{\LANL}{Los Alamos National Laboratory (LANL), Los Alamos, NM, 87545, USA}
\newcommand{\Louisiana}{Louisiana State University, Baton Rouge, LA, 70803, USA}
\newcommand{\Manchester}{The University of Manchester, Manchester M13 9PL, United Kingdom}
\newcommand{\MIT}{Massachusetts Institute of Technology (MIT), Cambridge, MA, 02139, USA}
\newcommand{\Michigan}{University of Michigan, Ann Arbor, MI, 48109, USA}
\newcommand{\MSU}{Michigan State University, East Lansing, MI 48824, USA}
\newcommand{\Minnesota}{University of Minnesota, Minneapolis, MN, 55455, USA}
\newcommand{\Nankai}{Nankai University, Nankai District, Tianjin 300071, China}
\newcommand{\NMSU}{New Mexico State University (NMSU), Las Cruces, NM, 88003, USA}
\newcommand{\Oxford}{University of Oxford, Oxford OX1 3RH, United Kingdom}
\newcommand{\Pitt}{University of Pittsburgh, Pittsburgh, PA, 15260, USA}
\newcommand{\QMUL}{Queen Mary University of London, London E1 4NS, United Kingdom}
\newcommand{\Rutgers}{Rutgers University, Piscataway, NJ, 08854, USA}
\newcommand{\SLAC}{SLAC National Accelerator Laboratory, Menlo Park, CA, 94025, USA}
\newcommand{\SDSMT}{South Dakota School of Mines and Technology (SDSMT), Rapid City, SD, 57701, USA}
\newcommand{\Maine}{University of Southern Maine, Portland, ME, 04104, USA}
\newcommand{\Syracuse}{Syracuse University, Syracuse, NY, 13244, USA}
\newcommand{\TelAviv}{Tel Aviv University, Tel Aviv, Israel, 69978}
\newcommand{\UTA}{University of Texas, Arlington, TX, 76019, USA}
\newcommand{\Tufts}{Tufts University, Medford, MA, 02155, USA}
\newcommand{\VTech}{Center for Neutrino Physics, Virginia Tech, Blacksburg, VA, 24061, USA}
\newcommand{\Warwick}{University of Warwick, Coventry CV4 7AL, United Kingdom}

\affiliation{\ANL}
\affiliation{\Bern}
\affiliation{\BNL}
\affiliation{\UCSB}
\affiliation{\Cambridge}
\affiliation{\CIEMAT}
\affiliation{\Chicago}
\affiliation{\Cincinnati}
\affiliation{\CSU}
\affiliation{\Columbia}
\affiliation{\Edinburgh}
\affiliation{\FNAL}
\affiliation{\Granada}
\affiliation{\IIT}
\affiliation{\ICL}
\affiliation{\Indiana}
\affiliation{\KSU}
\affiliation{\Lancaster}
\affiliation{\LANL}
\affiliation{\Louisiana}
\affiliation{\Manchester}
\affiliation{\MIT}
\affiliation{\Michigan}
\affiliation{\MSU}
\affiliation{\Minnesota}
\affiliation{\Nankai}
\affiliation{\NMSU}
\affiliation{\Oxford}
\affiliation{\Pitt}
\affiliation{\QMUL}
\affiliation{\Rutgers}
\affiliation{\SLAC}
\affiliation{\SDSMT}
\affiliation{\Maine}
\affiliation{\Syracuse}
\affiliation{\TelAviv}
\affiliation{\UTA}
\affiliation{\Tufts}
\affiliation{\VTech}
\affiliation{\Warwick}

\author{P.~Abratenko} \affiliation{\Tufts}
\author{D.~Andrade~Aldana} \affiliation{\IIT}
\author{L.~Arellano} \affiliation{\Manchester}
\author{J.~Asaadi} \affiliation{\UTA}
\author{A.~Ashkenazi}\affiliation{\TelAviv}
\author{S.~Balasubramanian}\affiliation{\FNAL}
\author{B.~Baller} \affiliation{\FNAL}
\author{A.~Barnard} \affiliation{\Oxford}
\author{G.~Barr} \affiliation{\Oxford}
\author{D.~Barrow} \affiliation{\Oxford}
% ...

\collaboration{The MicroBooNE Collaboration}
\thanks{microboone\_info@fnal.gov}\noaffiliation

\date{\today}%

\begin{abstract}
We present the strongest experimental limits to date on the mixing angle, $\theta$, with which a new scalar particle, $S$, mixes with the Higgs field in the mass range $\unit[110]{MeV}<m_S<\unit[155]{MeV}$.
This result uses the MicroBooNE liquid argon time projection chamber to search for decays of these Higgs-portal scalar particles through the $S\rightarrow e^+e^-$ channel with the decays of kaons in the NuMI neutrino beam acting as the source of the scalar particles.
The analysis uses an exposure of $2.01\times 10^{21}$ protons on target of NuMI beam data including periods when the beam focusing system was configured to focus positively charged hadrons and separate periods when negatively charged hadrons were focused.
The analysis searches for scalar particles produced from kaons decaying in flight in the beam's decay volume and at rest in the target and absorber.
At $m_S=\unit[125]{MeV}$ ($m_S=\unit[150]{MeV})$ we set a limit of $\theta<3.19\times 10^{-4}$ ($\theta<2.79\times 10^{-4}$) at the 95\% confidence level.
\end{abstract}

\maketitle

\section{\label{sec:Introduction}Introduction}

The exceptional imaging and particle-identification capabilities of liquid-argon time projection chambers (LArTPCs)
enable a broad program of searches for physics beyond the Standard Model (SM)~\cite{ref:DavideHNL,ref:uBEE,ref:uBMuMu,ref:MarsdenHNL,ref:Trident, ArgoNeuTHeavyAxion,ICARUS_GranSasso_DM,uBooNE_LightSterile2023,ArgoNeuT_MilliCharge,ArgoNeuT_tau_HNL}.
The existence of dark matter (DM)~\cite{ref:Planck,ref:DarkMatterReview} motivates many models of new physics. With the absence of direct observations of weakly interacting massive particles~\cite{ref:Xenon,ref:PandaX-II,ref:LUX} as dark matter candidates, models are being developed that propose additional sectors of DM particles that couple to the SM through new forces~\cite{ref:SecludedDarkMatter,ref:DarkSectorWorkshop,ref:FeebleReview}.
In this paper, we use the MicroBooNE detector~\cite{ref:MicroBooNEDetector}, exposed to the Neutrinos from the Main Injector (NuMI) beam~\cite{ref:NuMIBeam}, to test a model in which the Higgs field acts as a portal to a new sector of particles~\cite{ref:TheModelWeAreTesting,ref:HiggsPortalPhenomenology}.

The Higgs-portal model predicts the existence of an invisible sector of particles, which can act as DM candidates.
This invisible sector couples to the SM only through the Higgs terms in the SM Lagrangian. This coupling proceeds through a new scalar singlet state $S$, which mixes with the SM Higgs boson through a mixing angle $\theta$. The mixing angle $\theta$ and the mass of the scalar, $m_S$, are free parameters of the model. 

\begin{figure}
    \centering
    \includegraphics[width=\columnwidth]{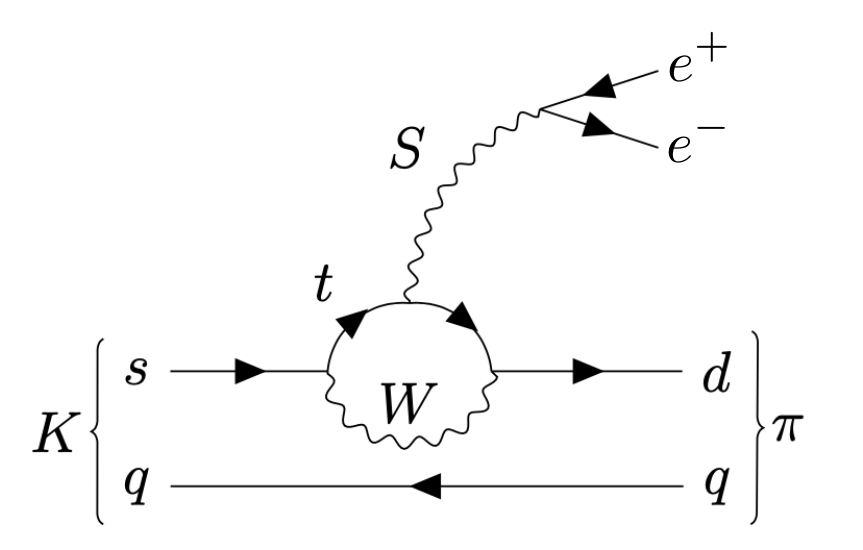}
    \caption{The dominant production channel for Higgs-portal scalar particles, $S$, is from kaon decays in the NuMI beam~\cite{ref:HiggsPortalPhenomenology}. The $e^+e^-$ decay mode of $S$ is the search mode of interest for this study. Here $q=u,d$ as both charged and neutral kaons contribute to production, and $u,d,s,t$ refer to the up, down, strange and top quarks.}
    \label{fig:FeynmanDiagram}
\end{figure}

We assume that the scalar particle $S$ is produced in the decays of $K$ mesons in the NuMI beam before it then decays in the MicroBooNE detector (Fig.~\ref{fig:FeynmanDiagram}).
Both the branching ratio for the production of the $S$ and the lifetime of the $S$ are proportional to $\theta^2$~\cite{ref:HiggsPortalPhenomenology}.
In this analysis, we search for the decay $S\rightarrow e^+e^-$. 
Since the $S$ would be produced in $K\to\pi S$ decays, the kinematic limits for this decay channel are $2m_e<m_S< m_K-m_\pi$. In this analysis, we restrict our search range to
$100$~MeV$<m_s<211$~MeV, since the E949~\cite{ref:E949} and NA69~\cite{ref:NA62} collaborations have published strong exclusions for lower masses, and the dominant decay mode in the range $m_S>2m_\mu$ is $S\to \mu^+\mu^-$, which leaves the $e^+e^-$ decay mode less sensitive.

\begin{figure*}[t!]
    \centering
    \includegraphics[width=\textwidth]{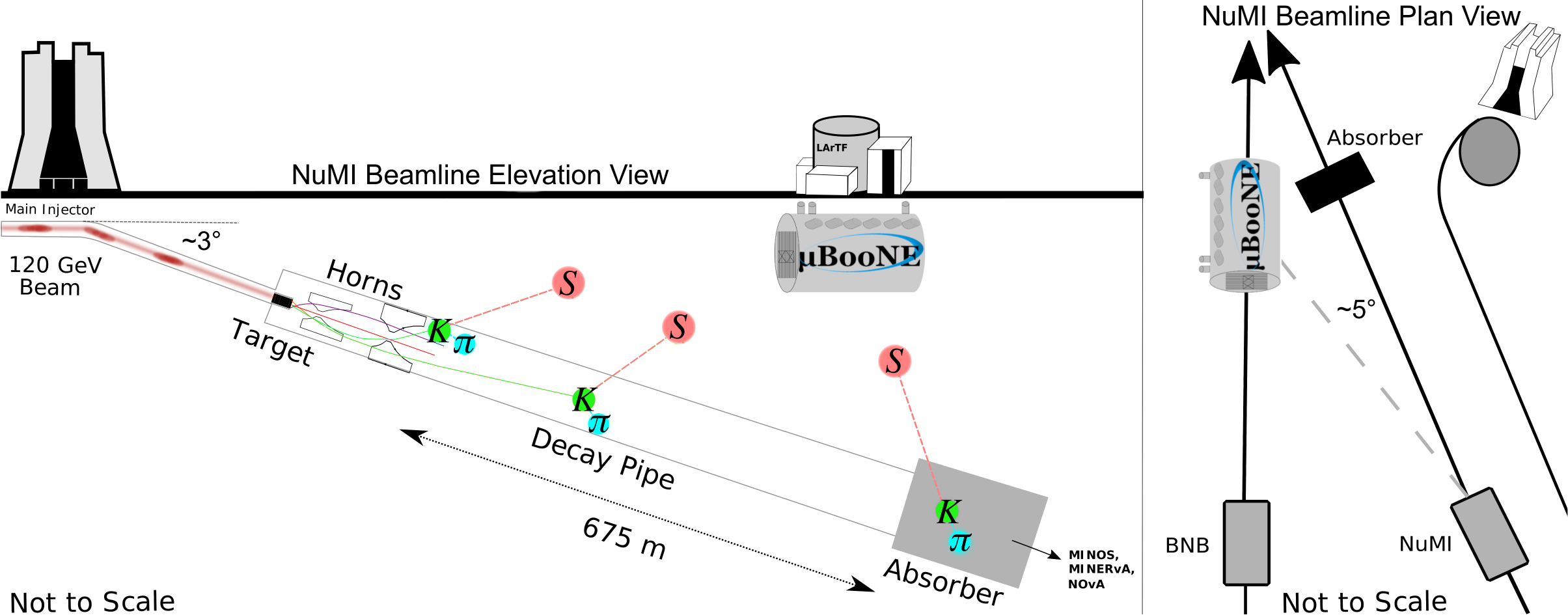}
    \caption{A schematic of the NuMI beamline, showing the location of the MicroBooNE detector with respect to the NuMI target, decay pipe, and hadron absorber. The schematic on the left shows an elevation view; the right schematic shows a plan view. Figure adapted from Ref.~\cite{ref:uBMuMu}.}
    \label{fig:NuMIBeam}
\end{figure*}

\section{MicroBooNE in the NuMI beam}

The MicroBooNE LArTPC~\cite{ref:MicroBooNEDetector} has an active volume of $\unit[2.3\times2.6\times10.4]{m^3}$. We define a fiducial volume at a distance of \unit[10]{cm} from each edge of the active volume and only analyze decay vertices within this fiducial volume. We use the \texttt{Pandora} pattern recognition package~\cite{ref:Pandora} to reconstruct candidates coincident in time with a NuMI beam spill that passes a trigger selection based on the detector's optical system. During MicroBooNE’s third running period (Run 3), a cosmic ray tagger (CRT) system of scintillator panels surrounding the LArTPC was commissioned, and its data was used to aid in vetoing cosmic muons.

A schematic of the NuMI beamline~\cite{ref:NuMIBeam}, located at the Fermi National Accelerator Laboratory (Fermilab), is shown in Figure ~\ref{fig:NuMIBeam}. Protons with an energy of \unit[120]{GeV} are incident on a graphite target resulting in the production of hadrons. Two magnetic horns focus the charged hadrons along a \unit[675]{m} long helium-filled decay pipe. 
Hadrons that reach the end of the decay pipe before decaying are stopped by an aluminium, steel, and concrete absorber of \unit[5.6]{m} height, \unit[5.5]{m} width, and \unit[8.5]{m} depth. This analysis uses $\unit[\fhcPot]{protons}$-on-target (POT) of exposure in which the NuMI focusing horns were configured to focus positively charged hadrons along the decay pipe (FHC), as well as $\unit[\rhcPot]{POT}$ of data in which the NuMI focusing horns were configured to focus negatively charged hadrons along the decay pipe (RHC).

\begin{figure*}
    \centering
    \includegraphics[width=0.8\textwidth]{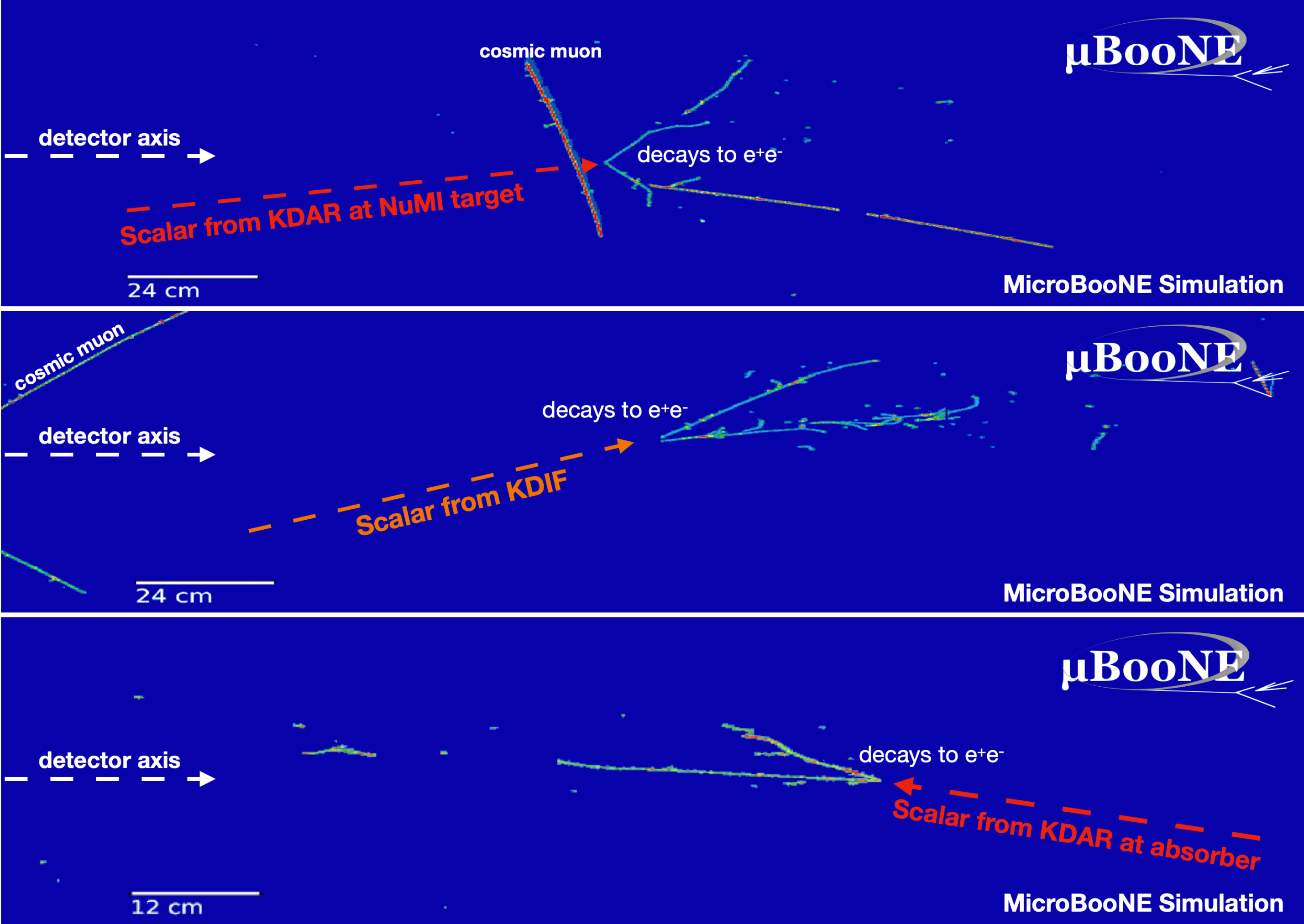}
    \caption{Three simulated decays of Higgs-portal scalar particles with masses of $\unit[200]{MeV}$ into $e^+e^-$ pairs. Top: a scalar produced from kaon decay at rest in the NuMI target. Middle: a scalar produced from kaon decay in flight in the NuMI decay pipe. Bottom: a scalar produced from kaon decay at rest in the NuMI hadron absorber.}
    \label{fig:EventDisplays}
\end{figure*}

\begin{figure}
    \centering
    \includegraphics[width=\columnwidth]{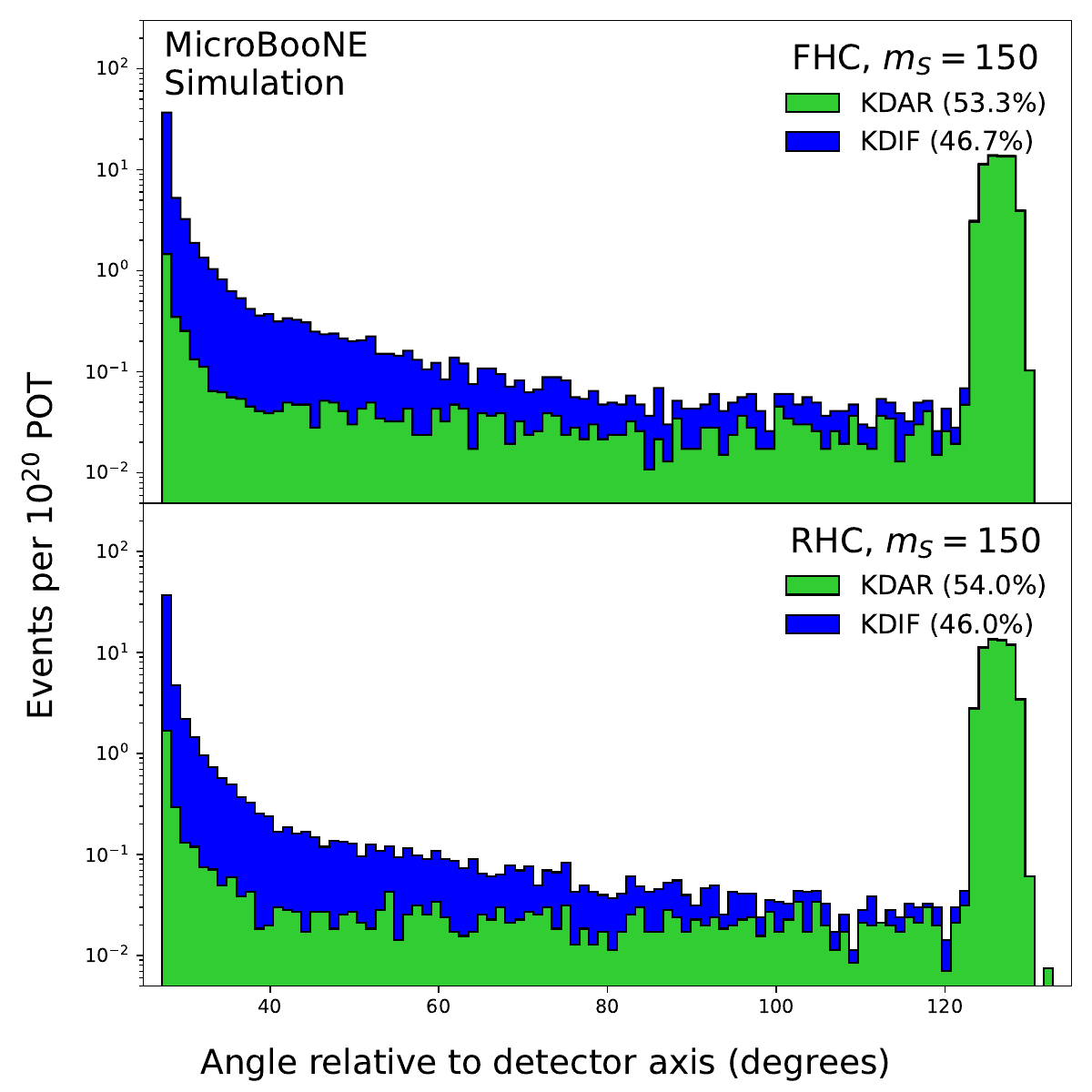}
    \caption{The production mechanism of simulated Higgs-portal scalar particles ($m_S=\unit[150]{MeV}$) for (a) Forward Horn Current (FHC) in which the NuMI horns are configured to focus positively charged mesons; (b) Reverse Horn Current (RHC), in which the NuMI horns are configured to focus negatively charged mesons.}
    \label{fig:ScalarParentage}
           \vspace{-95mm}
 \makebox[\textwidth][l]{\hspace{6.5cm}(a)} \\ 
 \vspace{34mm}
\makebox[\textwidth][l]{\hspace{6.5cm}(b)} \\ 
 \vspace{50mm}
\end{figure}

Figure~\ref{fig:EventDisplays} shows event displays of the simulated decays in the MicroBooNE detector of scalar particles produced in three locations in the NuMI beamline. Depending on their production point along the beamline, these particles will enter the detector from different directions. 
Scalar particles from a kaon decay at rest (KDAR) in the NuMI target will arrive at an angle of $8^\circ$, and those originating from KDAR in the NuMI absorber will arrive at $\sim120^\circ$. Particles from kaon decay in flight (KDIF) in the decay region downstream of the target will arrive at an angle between $8^\circ$ and $\sim120^\circ$.

KDIF events are concentrated towards smaller angles because the charged mesons are focused down the NuMI beamline, so scalars reaching MicroBooNE are predominantly from less off-axis kaons. Scalar production by KDAR is spherically symmetric, meaning a large number of scalars originate from the nearby absorber, located a distance of \unit[100]{m} from the detector, whereas a smaller number of scalars originate from the NuMI target, located a distance of \unit[680]{m} from the detector.

Simulation of the NuMI beam flux at MicroBooNE has been described in detail elsewhere~\cite{ref:KrishansPaper}. 
The flux of neutrinos through the MicroBooNE detector, which forms the main background for this search, is predicted by a \texttt{GEANTv4.9.2}~\cite{ref:GEANT4,ref:GEANT4Updates} simulation of the NuMI beamline, adjusted by the \texttt{PPFX} package~\cite{ref:MinervaNuMIFluxPaper}.
Updates to the NuMI flux model are discussed in~\cite{ref:NuMIFluxNote}, including an update to \texttt{GEANTv4.10.4} and target hall geometry, but are not included in this work.
Simulation of the scalar particle flux uses the same \texttt{GEANT4} simulation as for the kaon parents, and kaon decays to scalar particles are simulated according to the kinematics as described in Ref.~\cite{ref:HiggsPortalPhenomenology}.

The rate of KDAR in the NuMI absorber uses the prediction from the MiniBooNE collaboration, and was chosen because this prediction is consistent with the MiniBooNE measured KDAR charged current $\nu_\mu$ rate~\cite{ref:AbsorberFluxMeasurement}. Between the MiniBooNE and MicroBooNE operating periods, the NuMI beam changed from a \unit[96]{cm} to \unit[120]{cm} graphite target, reducing the rate of KDAR at the absorber. To account for this, an additional scaling factor is introduced. 
Production of scalar particles with a mass of $\unit[150]{MeV}$ is shown for each horn current configuration in Figure ~\ref{fig:ScalarParentage}. In RHC mode, KDAR produces 54\% of the simulated flux of scalar particles, with KDAR producing the remaining 46\%, and similarly for FHC.

\section{\label{sec:Selection} Identifying scalar boson decays in the MicroBooNE detector}

\begin{figure*}
    \centering
    \includegraphics[width=0.9\columnwidth,trim={0 0cm 0 2.3cm},clip]{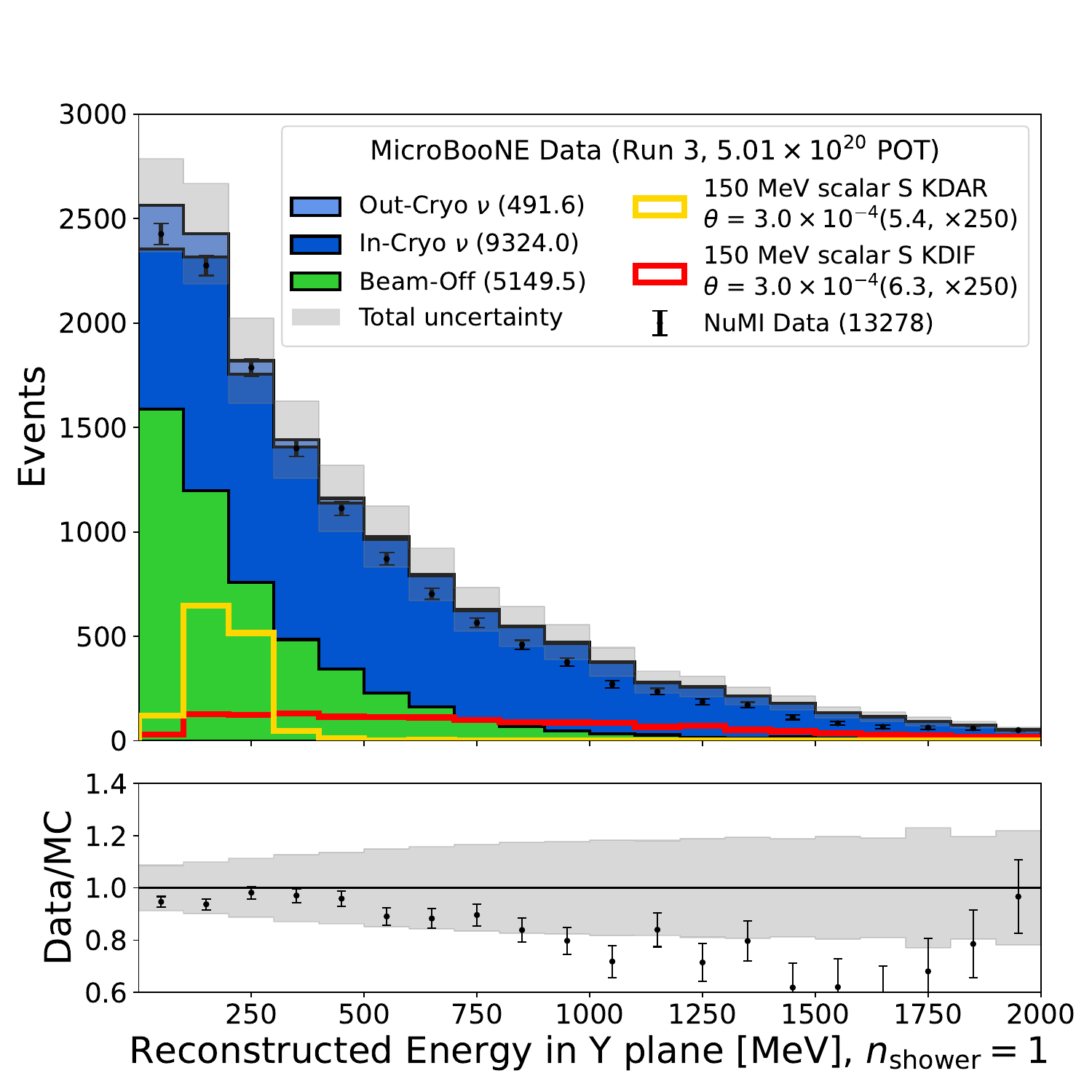}
    \includegraphics[width=0.9\columnwidth,trim={0 0cm 0 2.3cm},clip]{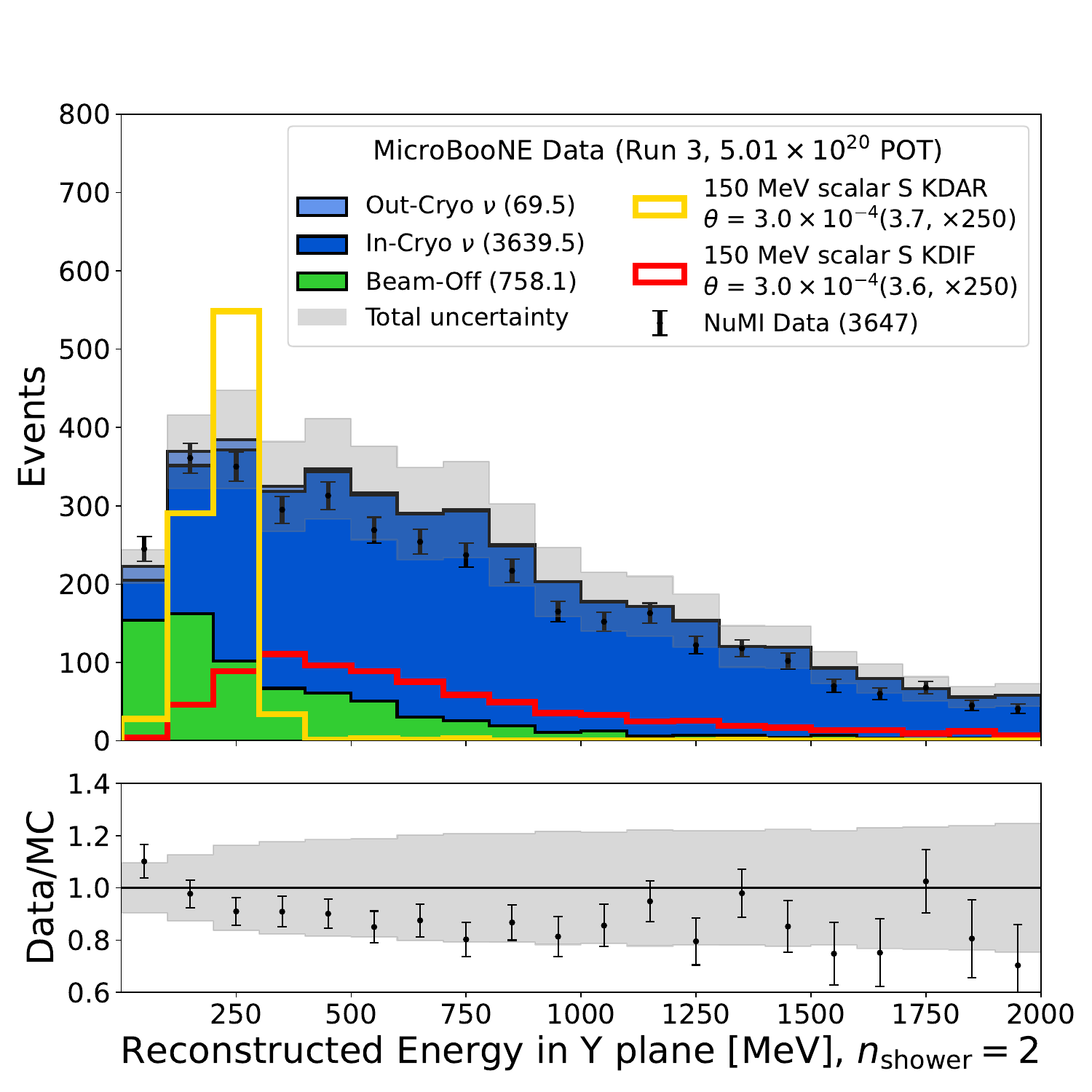}
    \includegraphics[width=0.9\columnwidth,trim={0 0cm 0 2.3cm},clip]{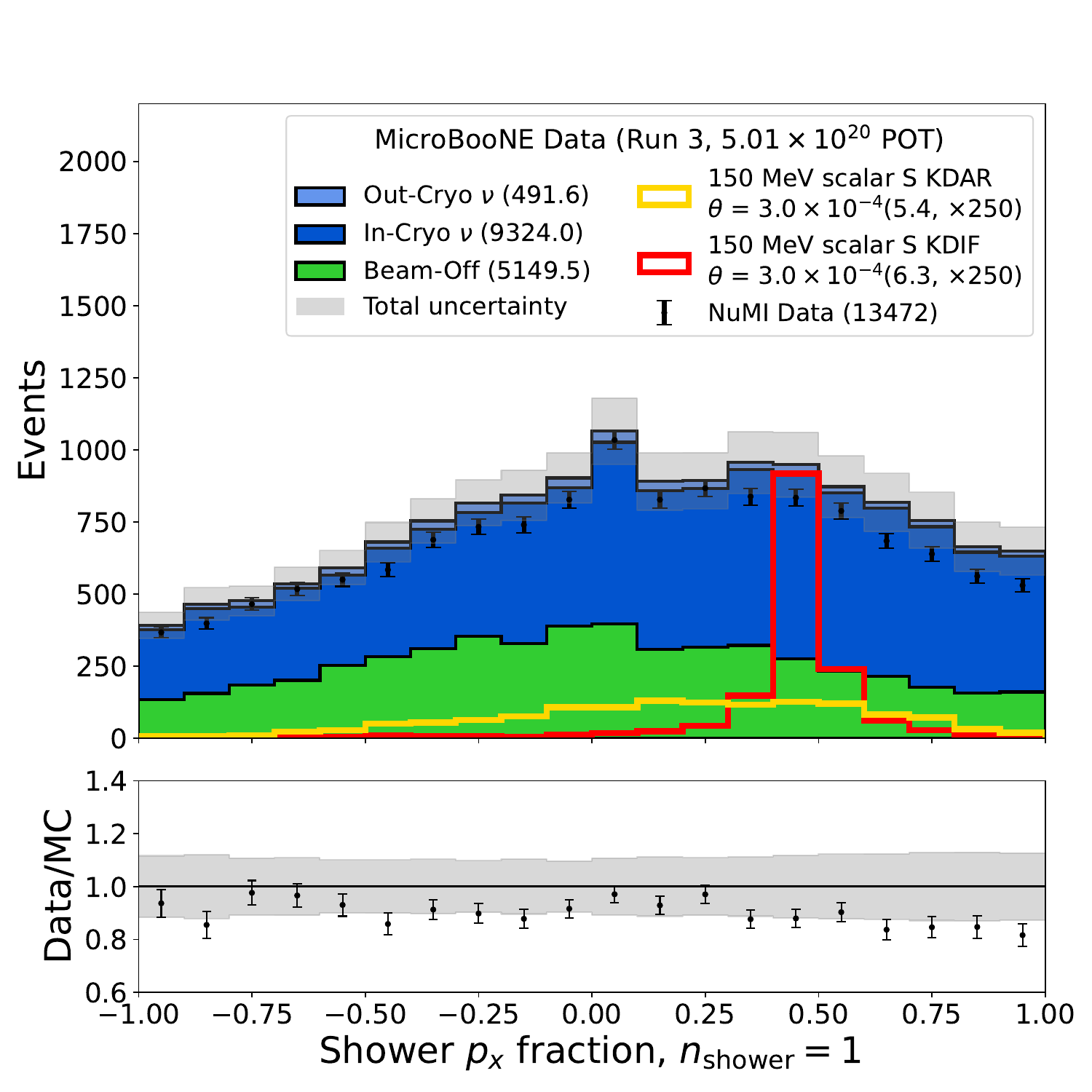}
    \includegraphics[width=0.9\columnwidth,trim={0 0cm 0 2.3cm},clip]{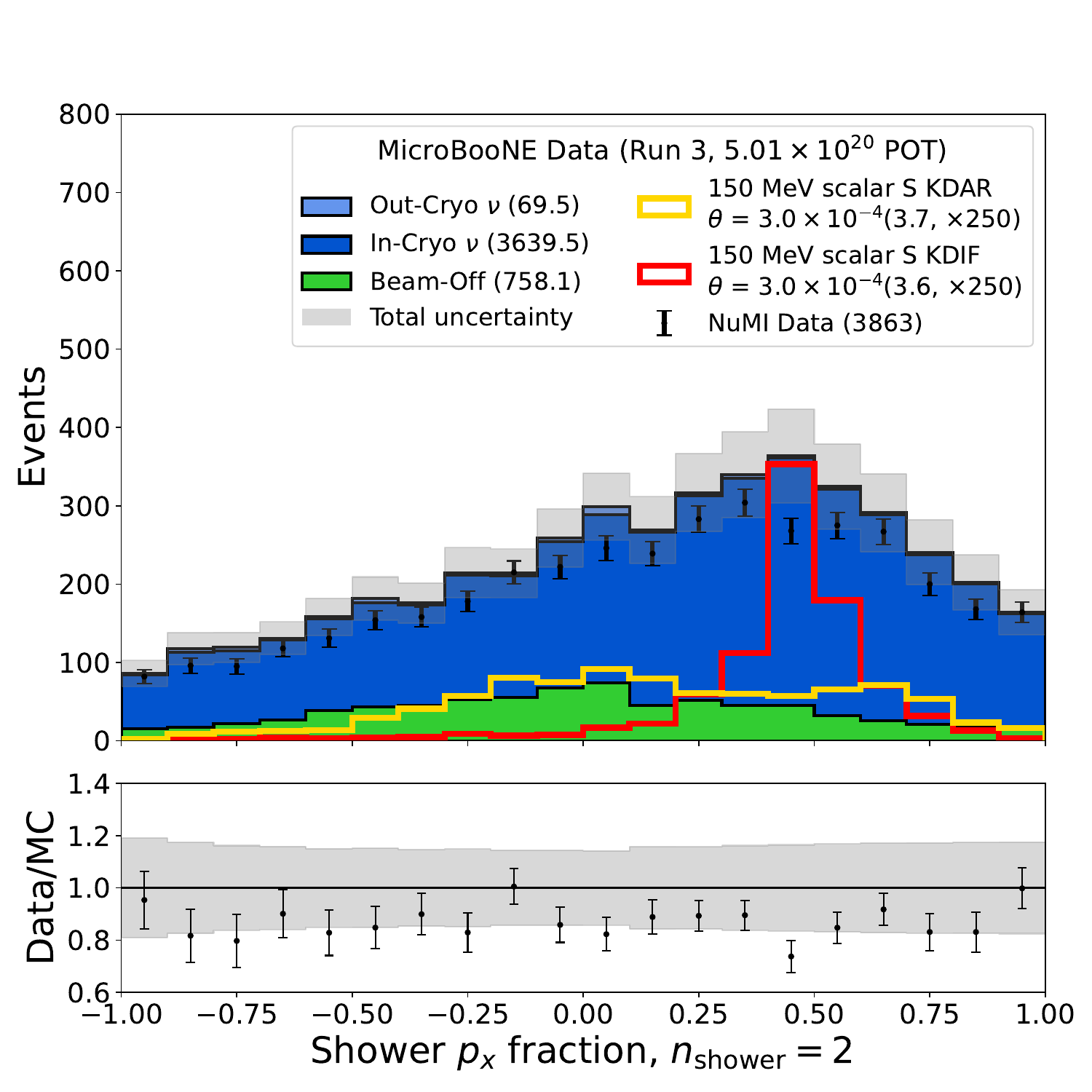}
    \includegraphics[width=0.9\columnwidth,trim={0 0cm 0 2.3cm},clip]{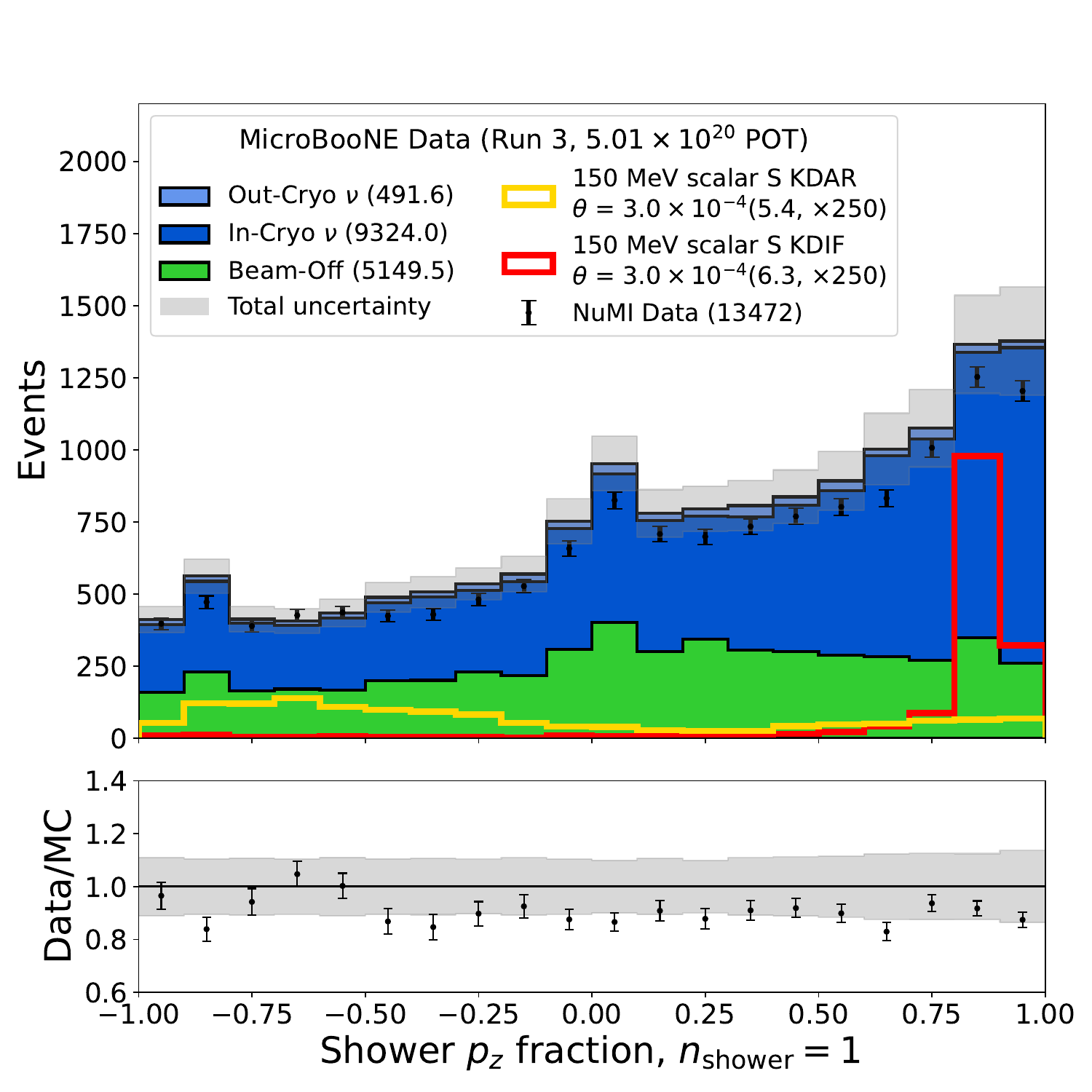}
    \includegraphics[width=0.9\columnwidth,trim={0 0cm 0 2.3cm},clip]{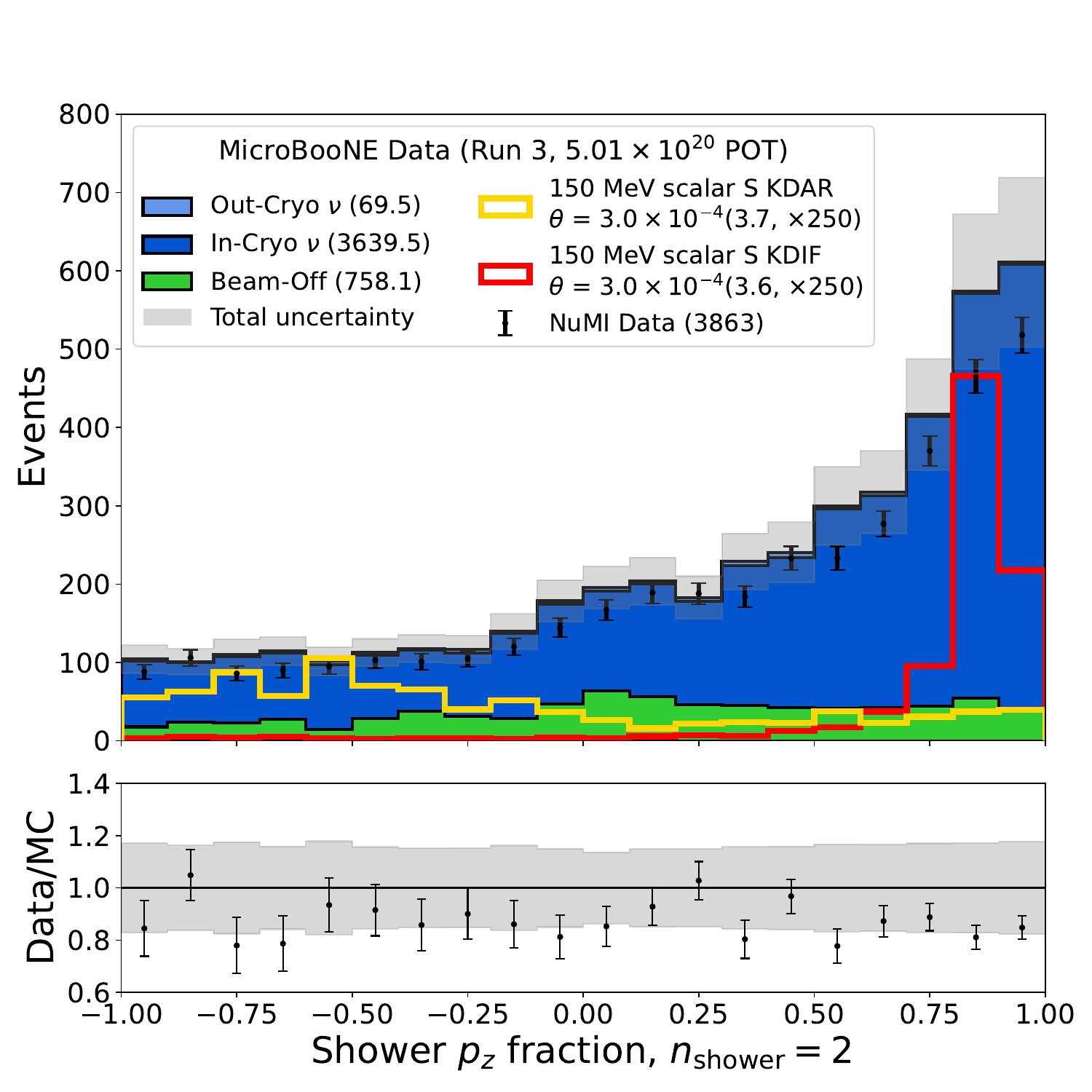}
    \caption{Three of the BDT input variables, shown for events in Run~3 trained for a scalar mass of $\unit[150]{MeV}$. (a) The reconstructed energy recorded in the TPC collection plane, the wires of which are oriented in the in the $y$-direction, for one-shower events. (b) The reconstructed energy recorded in the TPC collection plane for two-shower events. (c) The momentum fraction, in the $x$-direction, of the shower for one-shower events. (d) The momentum fraction, in the $x$-direction, of the highest-energy shower for two-shower events. (e) The momentum fraction, in the $z$-direction, of the shower for one-shower events. (f) The momentum fraction, in the $z$-direction, of the highest-energy shower for two-shower events. }
    \label{fig:BDTInputVariables}
           \vspace{-220mm}
 \makebox[\textwidth][l]{\hspace{7cm}(a)\hspace{7.5cm}(b)} \\ 
 \vspace{67mm}
\makebox[\textwidth][l]{\hspace{3cm}(c)\hspace{7.5cm}(d)} \\ 
 \vspace{67mm}
\makebox[\textwidth][l]{\hspace{3cm}(e)\hspace{7.5cm}(f)} \\ 
 \vspace{155mm}
\end{figure*}

The $e^+e^-$ topology should result in two shower-like clusters of activity, as shown in Fig.~\ref{fig:EventDisplays}.
However, fewer than half of simulated scalar boson decays result in two separate showers reconstructed by \texttt{Pandora}.
In a significant fraction of cases the two showers overlap and are reconstructed as a single shower.
We therefore divide our beam-like interactions into two samples: those with two reconstructed shower-like clusters and those with one reconstructed shower-like cluster.
Interactions with no reconstructed showers are rejected.

Neutrino interactions that mimic the $e^+e^-$ topology
are the main source of background. A further significant source of background are cosmic-ray muons passing through the detector. Due to the considerable cosmic flux, a number of muons mimic the single reconstructed shower signal topology, and to a lesser extent the two-shower topology. The \texttt{Pandora} pattern recognition package assigns a score to each candidate interaction that indicates whether it is cosmic-like or beam-like. We use this score, along with information from the CRT system when available, to isolate a sample of beam-like interactions.

To select signal-enhanced samples for analysis, a number of boosted decision trees (BDTs) are used.
These BDTs are trained on a full Monte Carlo simulation of the MicroBooNE detector with separate simulated samples of background neutrino interactions and signal Higgs-portal scalar decays.
Onto both of these simulated samples are overlaid data events taken with no beam passing through the detector in order to provide the simulations with realistic cosmic muons and detector noise.
The interactions of neutrinos in the detector are simulated with the \texttt{GENIEv3.0.6 G18\_10a\_02\_11a} ~\cite{ref:Genie,ref:Geniev3} generator modified with a custom tune of the interaction cross-sections on argon~\cite{ref:GenieTune}.
Particles exiting the simulated nucleus are propagated through a detector simulated with \texttt{GEANTv4.10.3.p03c}~\cite{ref:GEANT4,ref:GEANT4Updates}.
The decays of Higgs-portal scalar particles in the detector are simulated with custom-written code, and the particles resulting from the decay are, as for the case of neutrino interactions, propagated through the same detector simulation, implemented in LArSoft \cite{ref:LArSoft}.

We train separate BDTs to discriminate KDIF and KDAR topologies from background using information such as the direction of the interaction. For each scalar mass under investigation (given in Tab.~\ref{tab:EventCountsAndLimits}), eight separate BDTs are trained according to beam polarity and signal topology:
\begin{itemize}
    \item FHC, two-shower KDIF topology,
    \item FHC, two-shower KDAR topology,
    \item FHC, one-shower KDIF topology,
    \item FHC, one-shower KDAR topology,
    \item RHC, two-shower KDIF topology,
    \item RHC, two-shower KDAR topology,
    \item RHC, one-shower KDIF topology, and
    \item RHC, one-shower KDAR topology.
\end{itemize}
A pre-training is performed using a large number of variables to select 28 variables with the highest importance weights. Each BDT is then trained using these 28 variables, and the relative importance of the variables in each BDT will depend on the target topology.
Figure~\ref{fig:BDTInputVariables} shows the agreement between data and simulation for three variables that are typically the most important: the reconstructed energy recorded by the TPC collection plane, and the momentum fraction of the leading shower in the $x$ and $z$ directions, where $z$ is the direction of the detector axis, $y$ points vertically upwards, and $x$ points from the anode to the cathode of the TPC.
These three distributions show the distinguishing power between signal and background of these variables and illustrate the good agreement between data and simulation. The signal distributions are shown separately for scalar particles from KDAR and KDIF to illustrate why separate BDT trainings are performed to discriminate KDIF and KDAR topologies. The samples are primarily distinguished by the mono-energetic nature of scalars from KDAR and the differing incoming angle distributions. KDIF scalars are produced close to the target, whereas KDAR scalars are produced predominantly at the absorber.

\begin{figure*}
    \centering
    \includegraphics[width=\columnwidth]{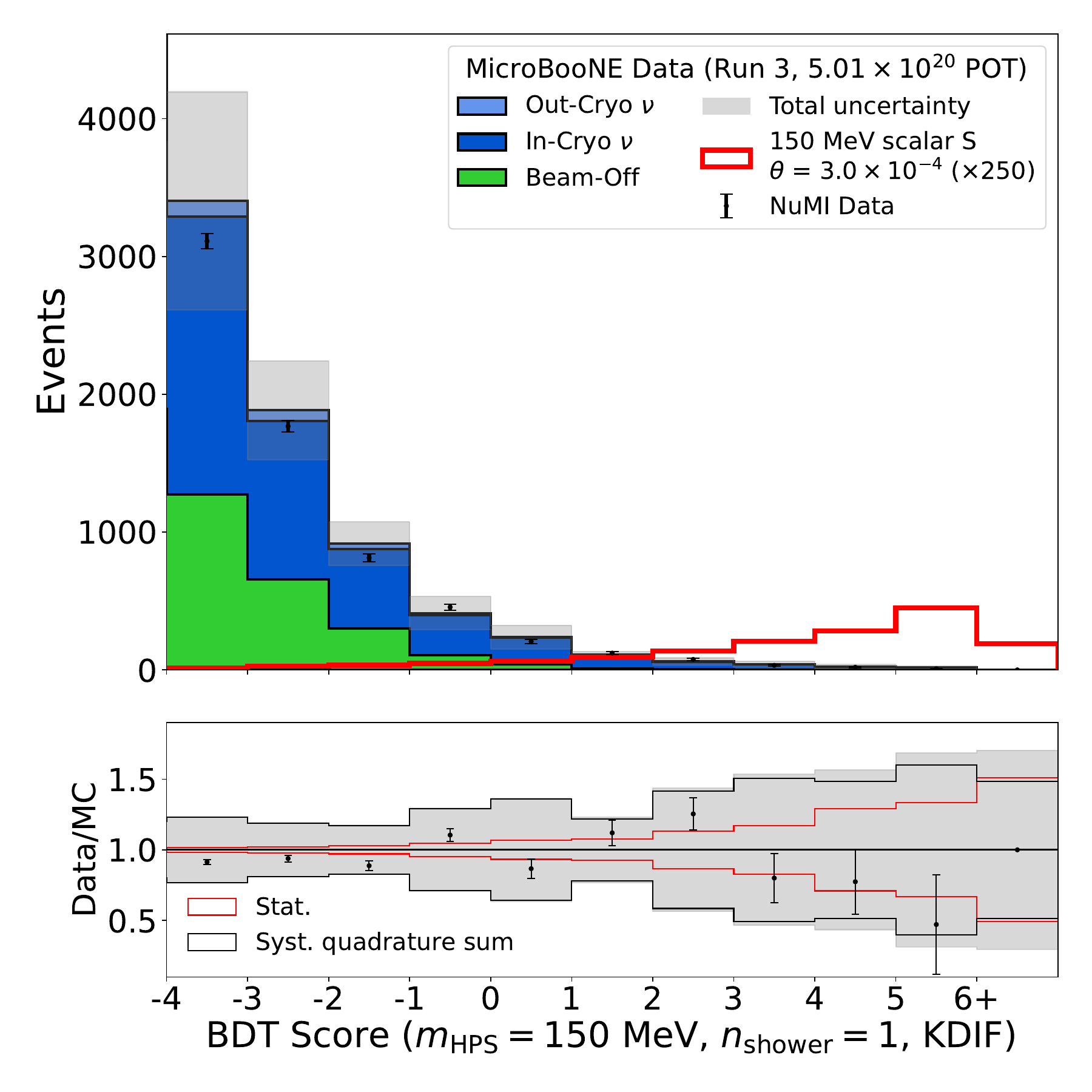}
    \includegraphics[width=\columnwidth]{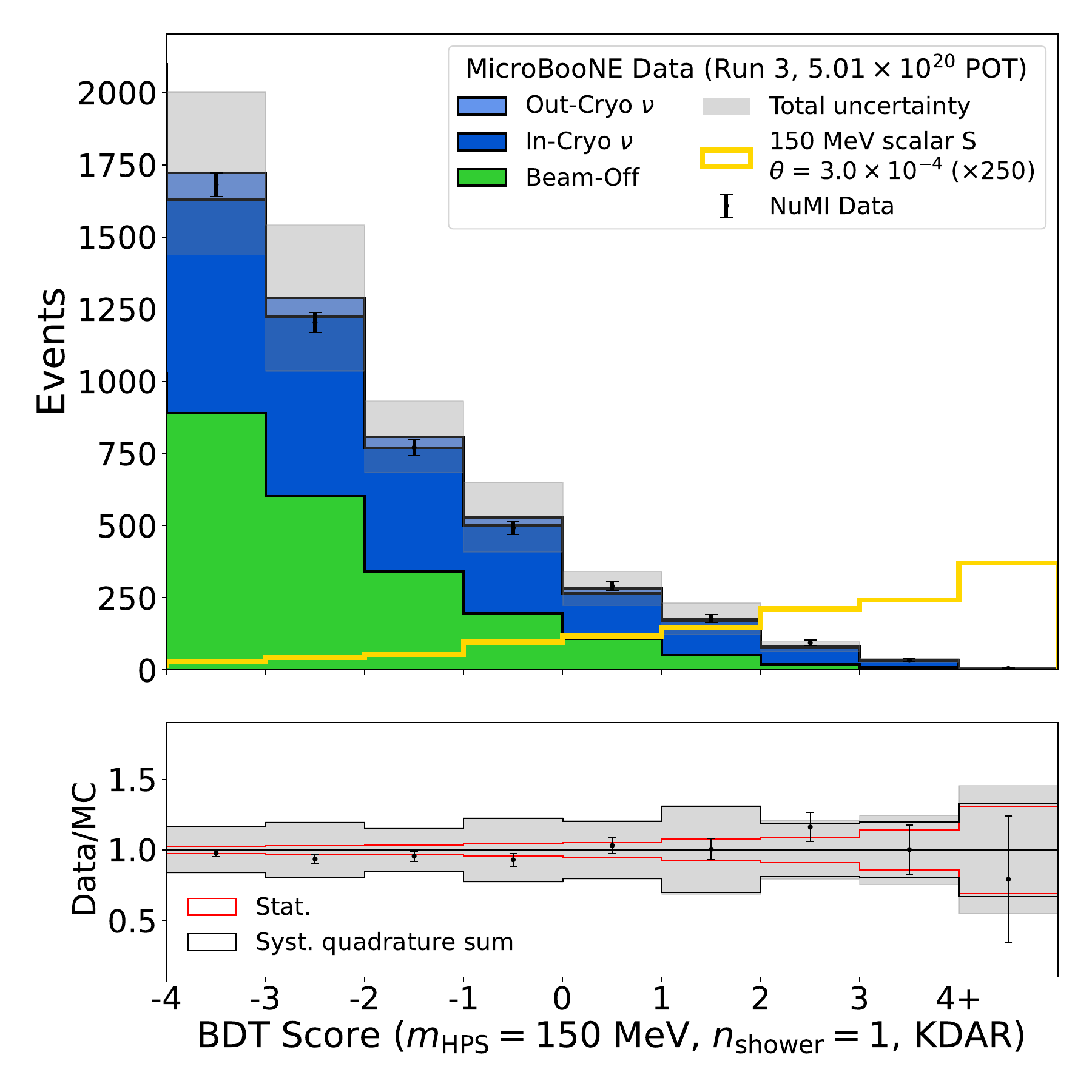}
    \includegraphics[width=\columnwidth]{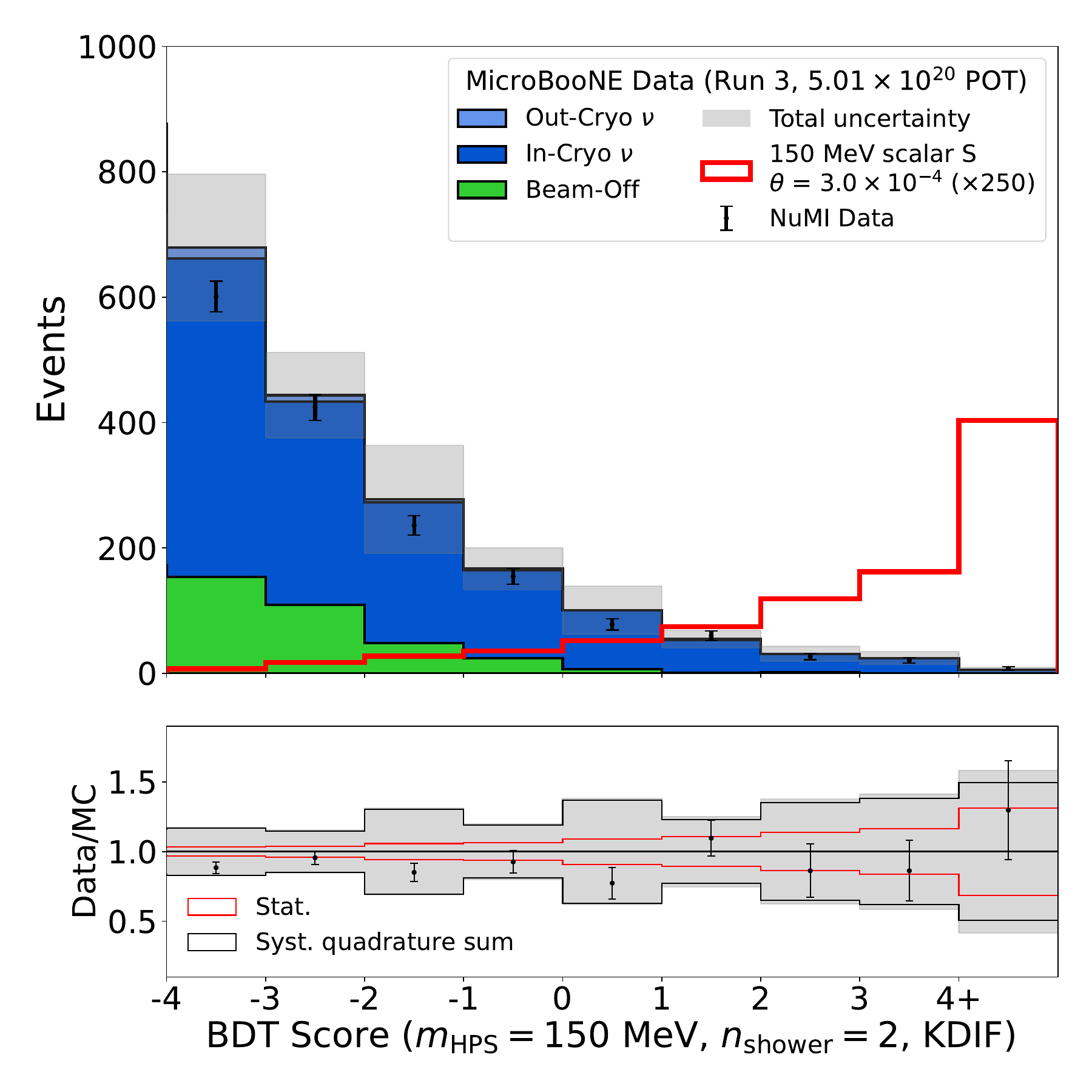}
    \includegraphics[width=\columnwidth]{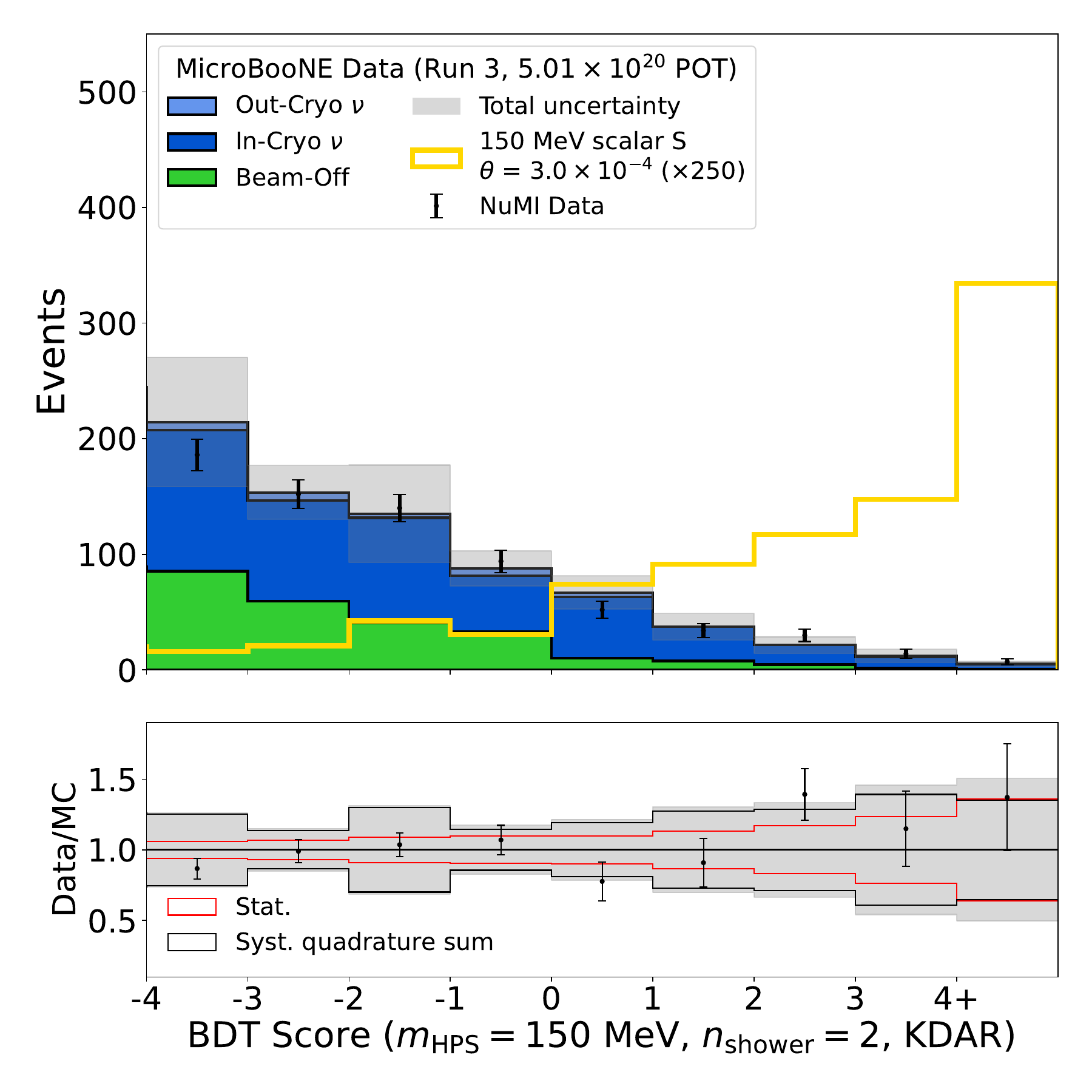}
    \caption {The distributions of BDT scores for the four BDTs trained to search for Higgs-portal scalar (HPS) particles with masses of $\unit[150]{MeV}$ in the Run 3 dataset.
    The quadrature sum of the uncertainties on the background prediction is shown by the gray band, where red indicates the MC statistical uncertainty, and black indicates systematic uncertainty.
    The simulated signal is shown at a normalization corresponding to $\theta = 3.0 \times 10^{-4}$, scaled up by a factor of 250. Overflow events are also included in the right-most bin of each BDT score distribution.
    (a) The BDT trained to distinguish KDIF topologies in the one-shower sample.
    (b) The BDT trained to distinguish KDAR topologies in the one-shower sample.
    (c) The BDT trained to distinguish KDIF topologies in the two-shower sample.
    (d) The BDT trained to distinguish KDAR topologies in the two-shower sample.}
    \label{fig:bdt_Run3_150}
           \vspace{-192mm}
 \makebox[\textwidth][l]{\hspace{3cm}(a)\hspace{8.4cm}(b)} \\ 
 \vspace{84mm}
\makebox[\textwidth][l]{\hspace{3cm}(c)\hspace{12cm}(d)} \\
 \vspace{155mm}
\end{figure*}

Figure~\ref{fig:bdt_Run3_150} shows the four BDTs trained for the search in the Run~3 sample at a scalar mass of $\unit[150]{MeV}$.
Good agreement is observed between the data and the background simulation across the full range of BDT values. The BDT can be seen to provide good separation between simulated signal and background events, particularly for the two-shower samples. 
The consistency between data and background prediction show no evidence of scalar decays within the MicroBooNE detector. Therefore this analysis will set limits on the Higgs-portal scalar parameter space.

In comparison to the previous MicroBooNE search for Higgs-portal scalar particles decaying into $e^+e^-$ pairs~\cite{ref:uBEE}, which used an exposure of $\unit[1.93\times10^{20}]{POT}$ of data from the NuMI beam in the Run~1 period, this analysis uses significantly more data. The previous search considered only mono-energetic scalar particles produced from KDAR in the NuMI absorber, whereas in this work we consider mono-energetic scalar particles produced from KDAR in both the NuMI target and absorber as well as scalar particles produced from KDIF in the NuMI decay volume, which will have a range of energies. 
The previous analysis was a purely topological search, while the study presented here uses significant additional information such as the reconstructed energy of the showers, and uses improved selection algorithms that rely on \texttt{Pandora}-based tools~\cite{ref:Pandora} to identify shower-like objects using the new BDTs described above to suppress background.

\section{Systematic uncertainties}

The dominant sources of systematic uncertainty at MicroBooNE arise from the modeling of the NuMI beam flux, neutrino background cross sections, modeling of the propagation of particles within the detector volume, and modeling of the detector response. The \texttt{PPFX} package~\cite{ref:MinervaNuMIFluxPaper}, used to predict the flux of hadrons in the NuMI beam, provides uncertainties on that flux arising from uncertainties on the production of hadrons in the NuMI target and the geometry of the beam focusing system. The full suite of uncertainties is applied to the simulated neutrino-induced background.

Uncertainties on the production of kaons impact the simulation of the Higgs-portal scalar signal. The \texttt{PPFX} package provides an energy-dependent uncertainty on kaons produced in the NuMI target. On charged kaons produced downstream of the target, a 40\% uncertainty is assigned based on model spreads in the absence of data. Approximately 30\% of the Higgs-portal scalar flux arises from $K^0_L$ decays. 

The \texttt{PPFX} prediction uses isospin symmetry to infer the neutral kaon flux based on charged kaon data~\cite{ref:IsospinSymmetryKaons} but assigns no uncertainty. To conservatively account for the systematic uncertainty introduced by the charged kaon rates and the uncertainty due to assumptions made in the flux calculation, an uncertainty of 100\%  is given for Higgs-portal scalar particles produced by $K^0_L$ decays. 
A 30\% uncertainty is assigned to the rate of KDAR in the NuMI absorber. This is derived from the 30\% model spread between the \texttt{MARS}~\cite{ref:MARS}, \texttt{FLUKA}~\cite{ref:Fluka}, and \texttt{GEANT4}~\cite{ref:GEANT4,ref:GEANT4Updates} simulations as described in~\cite{ref:AbsorberFluxMeasurement}.

Uncertainties on the neutrino-induced background arising from interaction cross-section modeling are described in~\cite{ref:GenieTune}.
Uncertainties on particle propagation through the detector are assessed using the \texttt{GEANT4Reweight} package~\cite{ref:GEANT4Reweight}.
Uncertainties on the response of the detector to ionization are simulated as described in~\cite{ref:DetVar}. 
Uncertainties on the detector's electric field map arising from a space charge effect and on charge recombination are separately accounted for.
The scintillation light-yield uncertainty is estimated by modeling  a $25\%$ decrease in light yield.
An uncertainty on the Rayleigh scattering length of argon is evaluated by varying this quantity from \unit[60]{cm} to \unit[90]{cm}.
Uncertainties on the attenuation length of scintillation light in argon are accounted for based on an observed decline in light levels over the course of detector operations. 
An uncertainty on the background arising from neutrino interactions outside the active detector region is also incorporated. All of the aforementioned systematic uncertainties are summed in quadrature, the total of which is shown in Fig.~\ref{fig:bdt_Run3_150} as a function of the BDT score.

\section{Results}

The limits on $\theta$ are obtained using the semi-frequentist CL$_\mathrm{S}$ method~\cite{ref:CLsJunk,ref:CLsRead} as implemented in the \texttt{pyhf} limit-setting software framework~\cite{ref:pyhf,ref:pyhf_joss,ref:AsymptoticLimitSetting}.
At each scalar-particle mass, the \texttt{pyhf} package uses the six highest-score bins from each of the eight BDT distributions, and scales the signal contributions simultaneously across these distributions to set limits on $\theta$.

\begin{figure}
    \centering
    \includegraphics[width=\columnwidth]{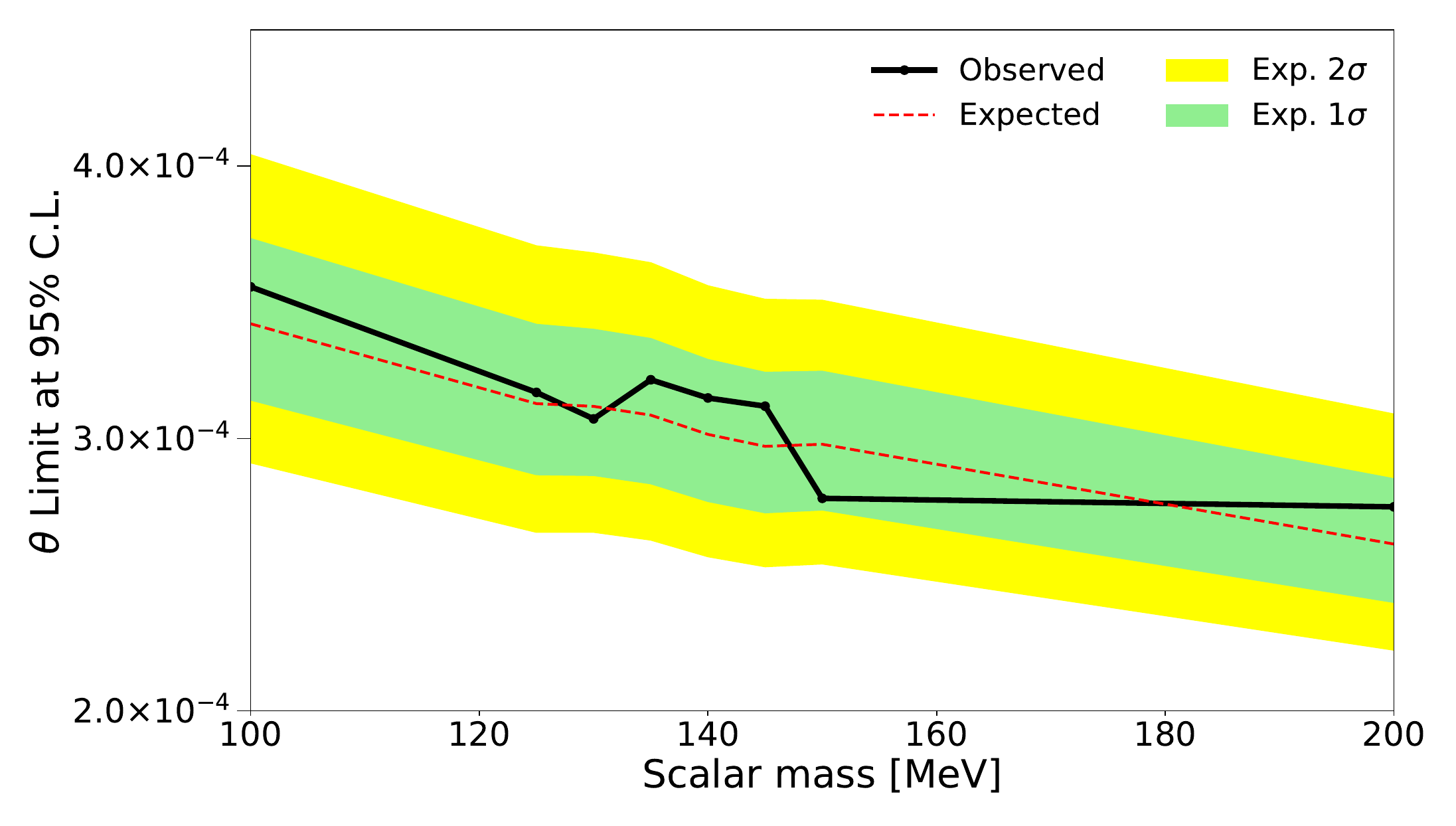}
    \caption{Limits set by this analysis on the mixing angle $\theta$ that governs the coupling of an invisible sector to the Higgs term in the SM Lagrangian. The limits are presented as a function of the mass, $m_S$, of the observable new scalar particle $S$ that would result from such a coupling, and are compared with the expected limits.}
    \label{fig:BrazilianResultsPlot}
\end{figure}

\begin{figure*}
    \centering
    \includegraphics[width=0.9\textwidth]{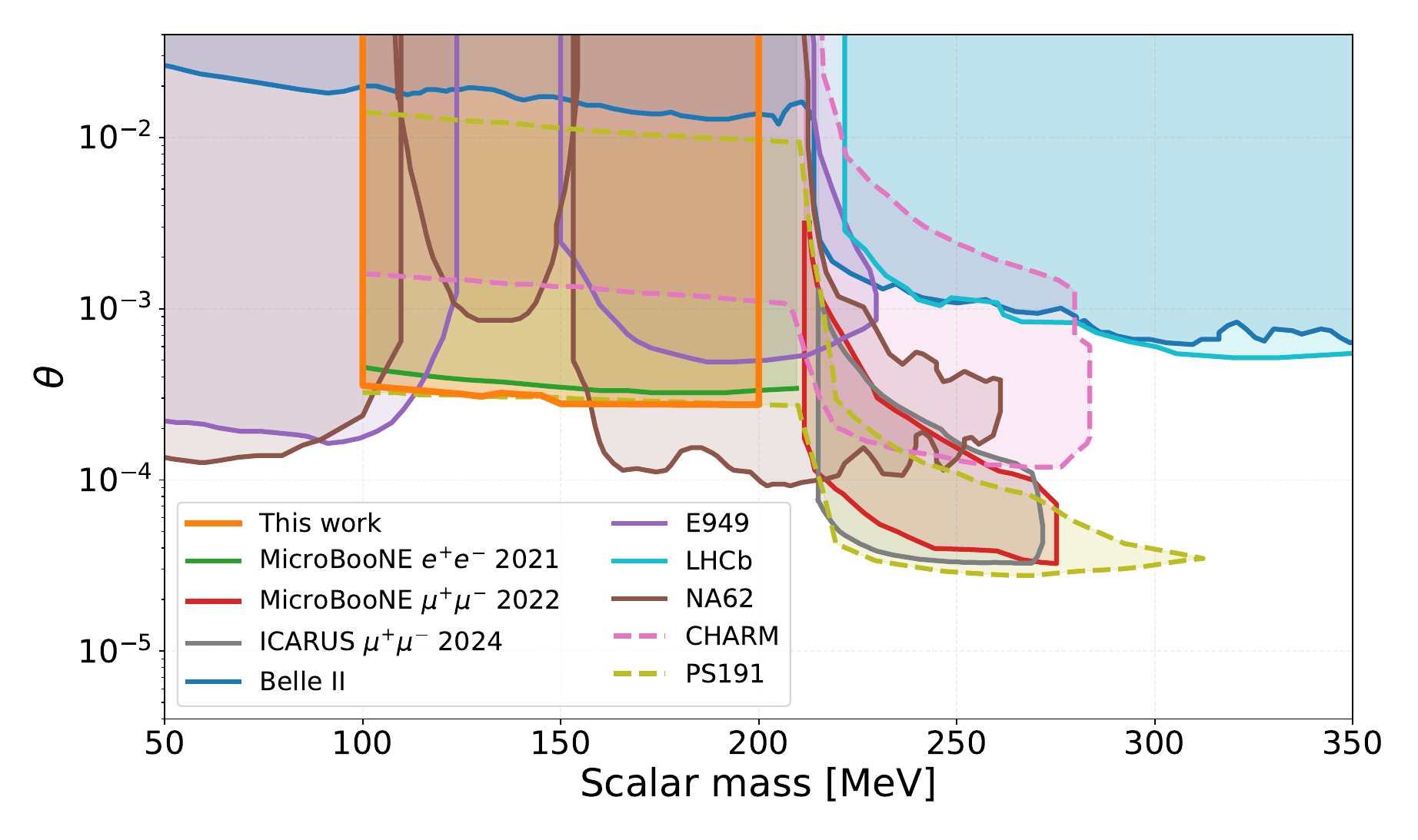}
    \caption{The 95\% C.L. limits on the mixing angle $\theta$ set by this analysis, as previously shown in Fig.~\ref{fig:BrazilianResultsPlot}, compared to limits set by previous measurements. Shown are direct limits set by the Belle-II~\cite{ref:BelleII}, E949~\cite{ref:E949}, LHCb~\cite{ref:LHCb1,ref:LHCb2}, NA62~\cite{ref:NA62} and ICARUS~\cite{ref:ICARUS_mumu} experiments, and previous limits from MicroBooNE~\cite{ref:uBEE,ref:uBMuMu}. Limits are also shown from reinterpretations of data from the CHARM~\cite{ref:CHARM} and PS191~\cite{ref:PS191} experiments. Limits from E949, NA62, PS191 and ICARUS, and the MicroBooNE $\mu^+\mu^-$ limit, are at 90\% C.L. All other limits are at 95\% C.L.}
    \label{fig:ResultsPlotWithOthers}
\end{figure*}

These limits are shown in Figure~\ref{fig:BrazilianResultsPlot} and Table~\ref{tab:EventCountsAndLimits}: when compared to the expected limits, they are consistent with the $\pm1\sigma$ expectation.
Table~\ref{tab:EventCountsAndLimits} also shows the total number of expected and observed events at each mass point, obtained by summing the highest-scoring bins across all BDTs.
Figure~\ref{fig:ResultsPlotWithOthers} compares the limits set in this work to pre-existing limits from other experiments. We set the strongest experimental limits to date on $\theta$ in the mass range $\lowMass<m_S<\highMass$: this is the region around the $\pi^0$ mass, in which the NA62~\cite{ref:NA62} and E949~\cite{ref:E949} limits are significantly weakened by $\pi^0$ backgrounds.
The previous strongest direct limit in this region is that set by a previous MicroBooNE search~\cite{ref:uBEE}.
A reinterpretation of PS191 data~\cite{ref:PS191} also presents a competitive limit to this work.

\begin{table}
    \centering

    \label{tab:EventCountsAndLimits}
    \begin{tabular}{ccrcr}
         \hline\hline 
         \rule{0pt}{3ex}$m_S$ & \multicolumn{2}{c}{95\% C.L. limit} &  \multicolumn{2}{c}{Selected events}\\
        $(\mathrm{MeV})$  & \multicolumn{2}{c}{on $\theta$ ($\times 10^{-4}$)} & \multicolumn{2}{c}{ (Highest score bins)} \\
        [1mm] \cline{2-3} \cline{4-5}
        \rule{0pt}{3ex} & Observed & Expected & Observed & Background \\[1mm] \hline
        \rule{0pt}{3ex}100 & 3.57 & $3.42^{+0.32}_{-0.28}$ & $95$ & $80.2^{+10.0}_{-10.0}$ \\
        \rule{0pt}{3ex}125 & 3.19 & $3.13^{+0.29}_{-0.26}$ & $93$ & $83.6^{+10.8}_{-10.8}$ \\
        \rule{0pt}{3ex}130 & 3.08 & $3.12^{+0.29}_{-0.25}$ & $92$& $96.6^{+11.5}_{-11.5}$\\
        \rule{0pt}{3ex}135 & 3.23 & $3.09^{+0.28}_{-0.25}$ & $114$ & $101^{+12}_{-12}$\\
        \rule{0pt}{3ex}140 & 3.16 & $3.02^{+0.28}_{-0.25}$ & $100$ & $80.4^{+9.5}_{-9.5}$ \\
        \rule{0pt}{3ex}145 & 3.14 & $2.97^{+0.27}_{-0.24}$ & $106$ & $87.1^{+9.9}_{-9.9}$ \\
        \rule{0pt}{3ex}150 & 2.79 & $2.98^{+0.27}_{-0.24}$ & $91$ & $95.1^{+10.3}_{-10.3}$ \\
        \rule{0pt}{3ex}200 & 2.76 & $2.61^{+0.24}_{-0.21}$ & $124$ & $112^{+15}_{-15}$ \\[1mm] 
         \hline\hline
    \end{tabular}
    \caption{Observed and expected limits on the mixing angle $\theta$, and the final observed and expected background events, given as a function of the mass, $m_S$, of the scalar particle $S$. }
    \label{tab:my_label}
\end{table}

Previous limits on the existence of Higgs-portal scalar particles have been set by the BELLE-II~\cite{ref:BelleII}, E949~\cite{ref:E949}, LHCb~\cite{ref:LHCb1,ref:LHCb2},  NA62~\cite{ref:NA62} and ICARUS~\cite{ref:ICARUS_mumu} experiments, and by the MicroBooNE experiment through the $e^+e^-$~\cite{ref:uBEE} and $\mu^+\mu^-$~\cite{ref:uBMuMu} decay channels. The previous MicroBooNE $e^+e^-$ result is the direct predecessor to this analysis. Additional constraints have been made by reinterpreting data from the CHARM~\cite{ref:CHARM} and PS191~\cite{ref:PS191} experiments. All these existing limits are shown in Fig.~\ref{fig:ResultsPlotWithOthers}. Additionally, not shown on the figure, some regions of small $m_S$ and small $\theta$ can be excluded from cosmological considerations~\cite{ref:Gordan}.

\section{Conclusions}

We have presented limits on the mixing angle $\theta$ that describes the coupling, through the Higgs mass term in the SM Lagrangian, of an invisible sector that could provide a source of dark matter particles. These limits have been set through a search for a new scalar singlet particle, $S$, that would be produced in kaon decays in the NuMI beam and be detected through its decays to $e^+e^-$ pairs in the MicroBooNE LArTPC.
This work uses kaons decaying in flight within the beam's decay volume, at rest in the target, and within the absorber. This approach significantly increases the expected scalar flux compared to previous limits set by MicroBooNE~\cite{ref:uBEE,ref:uBMuMu}. In this paper, we present the strongest experimental limits  to date on the mixing parameter $\theta$ in the mass range $\lowMass<m_S<\highMass$ corresponding to $\totalPot$ protons on target, the full NuMI beam dataset recorded by MicroBooNE.

\section*{Acknowledgements}

This document was prepared by the MicroBooNE collaboration using the resources of the Fermi National Accelerator Laboratory (Fermilab), a U.S. Department of Energy, Office of Science, HEP User Facility. Fermilab is managed by Fermi Research Alliance, LLC (FRA), acting under Contract No.\ DE-AC02-07CH11359. MicroBooNE is supported by the following: the U.S. Department of Energy, Office of Science, Offices of High Energy Physics and Nuclear Physics; the U.S. National Science Foundation; the Swiss National Science Foundation; the Science and Technology Facilities Council (STFC), part of the United Kingdom Research and Innovation; the Royal Society (United Kingdom); the UK Research and Innovation (UKRI) Future Leaders Fellowship; and The European Union’s Horizon 2020 Marie Sk\l{}odowska-Curie Actions. Additional support for the laser calibration system and cosmic ray tagger was provided by the Albert Einstein Center for Fundamental Physics, Bern, Switzerland. We also acknowledge the contributions of technical and scientific staff to the design, construction, and operation of the MicroBooNE detector as well as the contributions of past collaborators to the development of MicroBooNE analyses, without whom this work would not have been possible. For the purpose of open access, the authors have applied a Creative Commons Attribution (CC BY) licence to any Author Accepted Manuscript version arising from this submission.

\bibliography{bibfile}

\end{document}